\documentclass[twocolumn,prc,preprintnumbers,nofootinbib,superscriptaddress,showpacs,a4]{revtex4-2}
\usepackage{epsfig,graphicx,float,pst-all}
\usepackage{mathrsfs}
\usepackage[utf8]{inputenc}
\usepackage{rotating}
\usepackage{url}
\usepackage{esvect}
\usepackage{subfigure}
\usepackage{multirow}
\usepackage{amsmath}
\usepackage{amssymb}
\usepackage{amsfonts}
\usepackage{bm}
\usepackage{xcolor}
\usepackage{natbib}
\usepackage{hyperref}
\usepackage[warn]{textcomp}
\usepackage{gensymb}
\usepackage{footnote}
\usepackage{siunitx}
\usepackage{tabularx}
\usepackage{physics}
\usepackage{comment}
\usepackage{bm}
\usepackage{orcidlink}
\usepackage[normalem]{ulem}

\renewcommand{\vec}[1]{\boldsymbol{#1}}
\newcolumntype{C}[1]{>{\centering\arraybackslash}X}

\begin{document}
	\title{A two-cluster approach to the properties of one- and two-neutron-halo nuclei} %the possibility of
	\author{H. M. Maridi\orcidlink{0000-0002-2210-9897}}
	\email[Corresponding author: 
	%Permanent address: Department of Physics, Faculty of Applied Science, Taiz University, Taiz, Yemen; 
	]{hasan.maridi@manchester.ac.uk}
	\affiliation{Department of Physics and Astronomy, The University of Manchester, Manchester M13 9PL, UK}
	\author{Jagjit Singh\orcidlink{0000-0002-3198-4829}}
	%\email{jagjit.singh@manchester.ac.uk}
	\affiliation{Department of Physics and Astronomy, The University of Manchester, Manchester M13 9PL, UK}
	\affiliation{Department of Physics, Akal University, Talwandi Sabo, Bathinda, Punjab 151302, India}
	\affiliation{Research Centre for Nuclear Physics (RCNP), Osaka University, Ibaraki 567-0047, Japan}
	\author{N. R. Walet\orcidlink{0000-0002-2061-5534}}
	%\email{niels.walet@manchester.ac.uk}
	\affiliation{Department of Physics and Astronomy, The University of Manchester, Manchester M13 9PL, UK}
	\author{D. K. Sharp\orcidlink{0000-0003-1167-1603}}
	%\email{david.sharp@manchester.ac.uk}
	\affiliation{Department of Physics and Astronomy, The University of Manchester, Manchester M13 9PL, UK}
	\date{\today}

	\begin{abstract}
		In this work, we present a new approximate method for obtaining simple wave functions for the ground state of exotic nuclei with a neutron halo.
		We model the system as a two-cluster structure, treating the core and halo as inert objects. The relative wave function is expressed as a combination of simple harmonic oscillator states, with the oscillator parameter determined from the separation energy.
		Since these wave functions lack the expected exponential decay, we introduce a simple multiplicative renormalization factor based on the nuclear root mean square radius. This approach, combining oscillator wave functions and the renormalization factor, is then applied to calculate dipole strength distributions $dB(E1,\varepsilon)/d\varepsilon$ and Coulomb dissociation cross section $d\sigma(E1,\varepsilon)/d\varepsilon$ for several $1n$- and $2n$-halo nuclei. The results show excellent agreement with the available experimental data.
	\end{abstract}
		
	%\pacs{24.10.-i, 24.50.+g, 25.60.Tv}
	%\keywords{}

	\maketitle
	\section{\label{sec:int}Introduction}	
	The current generation of radioactive ion beam (RIB) facilities has opened up new opportunities for exploring the exotic features and responses of weakly-bound nuclei near the neutron ($n$) drip line \cite{Nowacki2021}. Such experiments offer a broad potential for advancing nuclear science, particularly in the understanding of shell evolution and formation of one- and two-neutron halos and skins in uncharted territories of the table of the nuclear elements. \cite{Tanihata2013,Torres2024}.
	
	The advances at these RIB facilities have led to the observation of $n$-halos in both light (\textit{e.g.}, $^6$He \cite{Aum99,Meister2002,Sun21,Fort2014,SFV16EPJ}, $^{11}$Li \cite{Nak06}, $^{11}$Be \cite{Nak94,Pal03,Fuk04,MORO2020}, $^{15,19}$C \cite{Nak09}, $^{17,19}$B \cite{Suzuki2002,Coo20}) and medium-mass (\textit{e.g.}, $^{22}$C \cite{Tanaka2010,TOGANO2016,JS19}, $^{29}$F \cite{Bagchi2020,Singh2020,Fortunato2020,Casal2020}, $^{37}$Mg \cite{Kobayashi2014}) nuclei. The future of the study of such halos looks promising, with potential $n$-halos predicted in various theoretical studies for nuclei such as $^{29}$Ne \cite{MDS21NPA}, $^{31}$F \cite{Masui2020,SSC22PRC}, $^{34}$Na \cite{SSC16PRC}, $^{39}$Na \cite{Zhang2023,SINGH2024}, $^{40}$Mg \cite{SINGH2024}, $^{42}$Al \cite{Zhang2023b}, many other isotopes of Mg, Al, Si, P, and S \cite{Li2024}, and even in heavier isotopes such as  $^{62,72}$Ca \cite{Hove2018,Horiuchi2022}. Such halo nuclei are characterized by their extended matter distributions \cite{Hansen1987,Zhu93}, strong di-neutron correlations \cite{Hagino2005,Kikuchi2016}, and large interaction cross-sections \cite{Tanihata1985}. Along with these, another notable feature of a halo nucleus is the presence of enhanced electric dipole ($E1$ ) strength at low energies. This can be observed via Coulomb dissociation experiments \cite{Aum13,Aumann2019}. 
	
	The enhanced low-lying $E1$ strength results from the similarity between weakly bound orbitals and low-lying continuum states near the threshold. This is in turn driven by the shallow binding and the nature of the single-particle states, rather than resonant behaviour. Theoretically, this feature can be reasonably accurately described as an inert core$+n$ for $1n$ \cite{RCPPNP,Moschini2019,Singh2021} halos, and core$+n+n$ for $2n$-halos \cite{Matsumoto2004,Casal2013,Fort2014,Sagawa2015,Singh2016,Horiuchi2006,JS19,Fortunato2020,Casal2020}. However, the ground state wave function for a $2n$-halo nucleus has an important core-dineutron component, that dominates the tail of the wave function\cite{Zhu93,Kee03}. Many other studies have used the dineutron model to describe the structure and dynamics of $^{6}$He \cite{Rus01,Kee03,Rus03a,Rus04,Rus05,Mac04,Mor07}.
	For $2n$-halo nuclei scattering from heavy-mass targets, the repulsive Coulomb force mainly excites the relative coordinate between the core and the pair of neutrons. This charge interaction pushes the charged core away from the target, while the nuclear interaction pulls the weakly-bound di-neutron closer. Hence, a simplified two-body (core-($2n$)) description of the three-body (core$+n+n$) projectile effectively captures the key reaction dynamics \cite{Mor07}. 
	Therefore, by employing the simplified two-cluster model, we aim to reasonably capture the essential dynamics of the scattering of halo nuclei of a heavy target, providing a practical approach to calculate low-lying dipole strength distributions and demonstrating its effectiveness through comparison with experimental data.
	
	The primary objective of this study is to develop a simple approximate two-cluster model (core+$n$ and core+$2n$), where we concentrate on determining simple wave functions for the ground state and associated effective (renormalized) operators. This method draws inspiration from the cluster orbital shell model approximation (COSM and COSMA) \cite{Zhu93, Suz88, Suz90, Suz91, Zhu93, Kor97a}. In this letter, we focus on four specific cases: $^{11}$Be and $^{15}$C (core+$n$ for $1n$-halos) and $^{6}$He and $^{11}$Li (core+$2n$ for $2n$-halos).
	Here, we calculate the low-lying dipole strength distribution using this new simple approximate approach. The results are then compared with experimental data, and we will show they provide accurate dipole-strength distributions.
	
	In the next section, we will give the formulation of our two-cluster model for $1n$- and $2n$-halos. In section 3  the resulting  $E1$-strength distribution for different $n$-halos are contrasted with experimental data. Finally, section 4 provides a summary of our research findings.
	
	%%%%%%%%%%%%%%%%%%%%%%%%%%%%%%%%%%%%%%%%%%%%%%%%%
	\section{\label{sec:formalism}Theory}
	\subsection{\label{sec:WF} Ground-state wave functions}
	In our two-cluster approach, the projectile is described as an inert core and a cluster of valence neutron(s). The reduced mass for their relative motion is $\mu$, and the orbital angular momentum for the relative motion of the core and valence neutron(s) is  $\vec{\ell}$. This is combined with the total internal angular momentum of the valence neutron(s) ($\vec{ S}$) to obtain the total relative angular momentum $\vec{ j}=\vec{ \ell}+\vec{ S}$. The total angular momentum of the projectile is then given by the coupling $\vec{J}=\vec{I}_c+\vec{ j}$, where $\vec{I}_c$ is the spin of the core. Finally, we use   $m_S$, $m_\ell$, $m_j$, and $M_c$ to denote the magnetic substates of the associated angular momenta, while $M$ is the projection of $J$. The wave functions can then be expressed in terms of the coordinate $\vec{r}$, the relative coordinate of the valence cluster relative to the c.m. of the core, as 
	\begin{equation} \label{eq:JM}
		\Psi_{JM}\left(  \vec{r}\right)  
		= \frac{u_{\ell j}^{J}\left( r\right)}{r}
		\sum_{m_j,\ M_{c}}  \left\langle jm_jI_{c}M_{c}|J{M}\right\rangle 
		%\ket{jm_j} \ket{I_{c}M_{c}} 
		\mathcal{Y}_{jm_j}^{\ell S}(\hat{\vec{r}}) \psi_{I_{c}M_{c}}, 
	\end{equation}
	with the spinor spherical harmonics
	\begin{equation}\label{eq:Yjm}
		\mathcal{Y}_{jm_j}^{\ell S}(\hat{\vec{r}})
		=\sum_{m_{\ell}\ ,\ m_{s}} \left\langle \ell m_{\ell} S m_{S}|j{m_j} \right\rangle Y_{\ell m_{\ell}}\left(\hat{\vec{r}}\right)
		\chi_{S m_{S}}\,.
	\end{equation}
	The function $u_{\ell j}^{J}$ is a radial wave function, while $\psi_{I_{c}M_{c}}$ and $\chi_{S m_{S}}$ denote the wave function of the core and the valence nucleons, respectively.
	
	In our current investigation, we assume there is no spin-orbit force, and use harmonic oscillator wave functions for the relative wave function of the valence cluster\cite{Bro60,Gol63,Mos69,Talmi1993}. The radial solutions is given as $u_{\ell j}^{J}\left( r\right)=rR_{n\ell} \left( r/b \right)$, 
	%		\begin{eqnarray}\label{eq:2}
		%		R_{n\ell} (r/b)&=& \left[ \frac{2^{\ell-n+2} (2\ell+2n+1)!!}{b^{2\ell+3}\sqrt{\pi} n!}\right]^{1/2} r^{\ell}e^{\frac{-r^2}{2 b^2}} \nonumber \\
		%		&& \sum_{k=0}^n \frac{(-1)^k 2^k n!}{k! (n-k)! (2\ell+2k+1)!!}  
		%		\left( \frac{r}{b}\right)^{2k},    
		%	\end{eqnarray}	
	with the oscillator parameter(length) $b$ given as
	\begin{equation}
		b \equiv b_{n\ell} = R_0/\sqrt{2n+\ell+\frac{3}{2}}.\label{eq:b}
	\end{equation} 
	The subtlety in this choice is that we use a different length parameter for each $n\ell$ pair, so that the wave functions all have the same radius $R_0$.
	
	In this study, we describe the valence neutrons in halo nuclei using $0s$, $0p$, $0d$, and $1s$ states. The $R_0$ value is expressed in terms of the binding energy of the valence neutron cluster to the core, $\mathcal{E}_0$, using an identity for harmonic oscillator states,
	\begin{equation}\label{eq:R0}
		R_0
		=\sqrt{\frac{3\hbar^2}{2\mu \mathcal{E}_0}}.
	\end{equation} 
	This relationship allows the determination of the free parameter $R_0$ in cases where experimental data for the core and subsystems are lacking, ensuring a reasonable and consistent approach across different scenarios.
%================================
	\subsection{\label{sec:dB} Dipole strength distributions} %($dB(EL)/d\varepsilon$)}
%================================
The reduced transition probability for an $E\lambda$ excitation between two states is normally presented as \cite{Typ05,Ber03}
\begin{equation}\label{eq:B}
	B\left(E\lambda;\ \ell_{0}j_{0}J_{0}\rightarrow ljJ\right)  =	\frac{\left\vert \left\langle \ell jJ\left\Vert \mathcal{O}_{E\lambda}\right\Vert \ell_0 j_0 J_{0}\right\rangle \right\vert ^{2}}{2J_{0}+1}.
\end{equation}
Here $\ket{\ell_0 j_0 J_{0}}$ represents the ground state and $\ket{\ell j J}$ stands for states with total angular momentum $J$ (for continuum states we label it with continuum energy $\varepsilon$ and momentum $\hbar k$). The quantity $\mathcal{O}_{E\lambda}$ is the electric multipole operator of order $\lambda$ and is given by $Z_{\rm eff}^{(\lambda)} r^{\lambda}Y_{\lambda\mu}\left(\hat{\vec{r}}\right)$, where we take the effective charge of the form
\begin{equation}\label{Zeff}
	Z_{\rm eff}^{(\lambda)}e=Z_{v}e\left(  -\frac{A_{c}}{A}\right)  ^{\lambda}%
	+Z_{c}e\left(  \frac{A_{v}}{A}\right)  ^{\lambda}
\end{equation}
where $Z, A$, $Z_c, A_c$ and $Z_v.A_v$  are the charge and mass number of the nucleus, the core and valence nucleon(s), respectively. 

For the electromagnetic breakup of the nucleus into the core and cluster of valence nucleon(s) with relative energy $\varepsilon$ and momentum $\hbar k$, the reduced transition probability can be given by summation over all possible angular momentum states of the continuum as \cite{Ber03}
\begin{align}\label{eq:dB}
	\frac{dB(E\lambda)}{d\varepsilon} &
	=\frac{\mu k}{(2\pi)^3 \hbar^2} \sum_{\ell,j,I_{c},J} \frac{\left\vert \left\langle k\ell jJ\left\Vert \mathcal{O}_{E\lambda
		}\right\Vert \ell_0 j_0 J_{0}\right\rangle \right\vert ^{2}}{2J_{0}+1}.
\end{align}
The initial and final radial wave functions $u_{l_{0}j_{0}}^{J_{0}}$ and $u_{\varepsilon lj}^{J}$ do not depend on the angular momentum $j$ and $J$ and for the summation over all possible $j$ and $J$, Eq.(\ref{eq:dB}) can be reduced to
\begin{eqnarray}\label{eq:dBl}
	\frac{dB(E\lambda)}{d\varepsilon} &=& \frac{(2\lambda+1)}{4 \pi}(Z_{\rm eff}^{(\lambda)}e)^{2} \sum_{l} \langle \ell_{0} 0\lambda 0|\ell 0 \rangle^{2} \nonumber\\ 
	&&\times \Bigg\vert   \int_{0}^{\infty}dr\
	r^{\lambda }\ u_{\varepsilon \ell j}^{J}\left(  r\right)  \
	u_{\ell_{0}j_{0}}^{J_{0}}\left(  r\right)\Bigg\vert^{2}. 
\end{eqnarray} 
We use the asymptotic expansion of the radial wave function, i.e., plane waves expanded in terms of spherical Bessel functions. Thus, Eq.~(\ref{eq:dBl}) gives us the low-lying multipole strength distribution for the valence neutron(s) in weakly bound systems. 

To include the different spins of the core, the initial bound wave function is given as a linear combination of these different core spins with valence nucleon(s) spin \cite{Nak23}
\begin{eqnarray} 
	%\Psi^{0}= 
	|J_0 M_0 \rangle = \sum_{n_0 \ell_0 j_0} \alpha(I_c^{\pi},n_0 \ell_0 j_0)|I_c^{\pi}\otimes n\ell_0 j_0\rangle  
\end{eqnarray}
in which the summation over the weights or the square of spectroscopic amplitudes, $\sum_{n_0 \ell_0 j_0} [\alpha(I_c^{\pi},n_0 \ell_0 j_0)]^2$ is unity,
%\begin{equation} \label{eq:A2}
%	\sum_{n_0 \ell_0 j_0} [A(I_c^{\pi},n_0 \ell_0 j_0)]^2=1
%\end{equation}
and then Eq.~(\ref{eq:dBl}) becomes
\begin{align}\label{eq:dBl2}
	\frac{dB(E\lambda)}{d\varepsilon} &=\frac{(2\lambda+1)}{4 \pi}(Z_{\rm eff}^{(\lambda)}e)^{2} \sum_{n_0 \ell_0 j_0} [\alpha(I_c^{\pi},n_0 \ell_0 j_0)]^2 \nonumber\\ &
	\times\sum_{\ell} \langle \ell_{0} 0\lambda 0|\ell 0 \rangle^{2} 
	\ \Bigg\vert   \int_{0}^{\infty}dr\
	r^{\lambda }\ u_{\varepsilon \ell j}^{J}\left(  r\right)  \
	u_{\ell_{0}j_{0}}^{J_{0}}\left(  r\right)\Bigg\vert^{2}. 
\end{align} 
For an $E1$ dipole transition from a bound state to all possible final states, the dipole response distribution is given by
	\begin{align}\label{eq:dB10}
		\frac{dB(E1)}{d\varepsilon} &=\frac{3}{4 \pi}(Z_{\rm eff}^{(1)}e)^{2} \sum_{n_0 \ell_0 j_0} [\alpha(I_c^{\pi},n_0 \ell_0 j_0)]^2 \nonumber\\ &
		\times\sum_{\ell} \langle \ell_{0} 01 0|\ell 0 \rangle^{2} 
		\ \Bigg\vert   \int_{0}^{\infty}dr\
		r\ u_{\varepsilon \ell j}^{J}\left(  r\right)  \
		u_{\ell_{0}j_{0}}^{J_{0}}\left(  r\right)\Bigg\vert^{2}. 
	\end{align} 
The total strength for the $E1$ transition is obtained by integrating over all energy states, 
	\begin{equation} \label{eq:Br1}
		B(E1) =\int_{0}^\infty \frac{dB(E1)}{d\varepsilon}d\varepsilon=\frac{3}{4\pi}(Z_{\rm eff}^{(1)}e)^{2}  \left\langle r_{cv}^2 \right\rangle  \,.
	\end{equation}
    For all harmonic oscillator wave functions, $R_{n\ell}$, the calculated value of $B(E1)$ is given by $\frac{3}{4\pi}(Z_{\rm eff}^{(1)}e)^{2} R_0^2$. However, due to the Gaussian tail of such wave functions, this expression tends to overestimate both the experimental values of $B(E1)$ and $\left\langle r_{cv}^2 \right\rangle$. To account for this discrepancy, we introduce a phenomenological multiplicative correction factor such that $\left\langle r_{cv}^2 \right\rangle_{exp}/R_0^2=C_0^2$, as discussed in Sec. \ref{sec:norm}.
%=======================================================

%\subsection{\label{sec:summary} Sum rule}
%=======================================================
The soft dipole mode (SDM) satisfies an energy-weighted sum rule (EWSR) which is evaluated as \cite{Fuk04}
\begin{equation} \label{eq:Sumrule1B}
	S_1=\int_{0}^\infty (\mathcal{E}_0+\varepsilon)\frac{dB(E1)}{d\varepsilon}d\varepsilon\,  .
\end{equation}
If the transition is purely single-paricle, this sum evaluates to \cite{Alh82,Suz91,Fuk04}
%\begin{align} \label{eq:Sumrule0}
%	S_1(E1;A_1+A_2)&=S_1(E1;A)-S_1(E1;A_1)-S_1(E1;A_2) \nonumber \\
%	&=\frac{9}{4\pi}\frac{\hbar^2 e^2}{2m}\left( \frac{N Z}{A}-\frac{N_1 Z_1}{A_1}-\frac{N_2 Z_2}{A_2}\right) \nonumber \\
%	&=\frac{9}{4\pi} \frac{(Z_1 A_2 - Z_2 A_1)^2}{A A_1 A_2}  \left( \frac{\hbar^2 e^2}{2m}\right)   \,,
%\end{align} 
\begin{equation} \label{eq:Sumrule1}
	%		S(E1,SDM)=
	S_1=\frac{9}{4\pi}\frac{\hbar^2 e^2}{2m}\left( \frac{N Z}{A}-\frac{N_c Z_c}{A_c}\right)    \,,
\end{equation}
where $N$ and $N_c$ are the neutron number of the nucleus and its core, respectively.
The ratio between the observed sum Eq.~(\ref{eq:Sumrule1B}) and the cluster sum Eq.~(\ref{eq:Sumrule1}) can also be used to extract the spectroscopic factor for the
halo state \cite{Fuk04}. 
%=======================================================
%\subsection{\label{sec:summary} cross sections}
%=======================================================

The Coulomb dissociation cross section for the electric dipole ($E1$) can be obtained from $dB(E1,\varepsilon)/d\varepsilon$ using the equivalent photon method \cite{Ber88}:
\begin{equation}
	\label{eq:dsde2}
	{d\sigma_{CD} \over d\varepsilon} = {16 \pi^{3} \over 9 \hbar c} {N_{E1}(\varepsilon_x)} {dB(E1,\varepsilon) \over d\varepsilon}, 
\end{equation} 
where $\varepsilon_x=\mathcal{E}_0+\varepsilon$ is the excitation energy and $N_{E1}(\varepsilon_x)$ is the number of virtual photons with energy $\varepsilon_x$.
\begin{comment}
	exchanged in a collision and given for high energies as \cite{Ber88}
	\begin{equation}
		\label{eq:ne}
		N_{E1}(\varepsilon_x) = {2 Z_T^2 \alpha \over \pi} \left( \frac{c}{v}\right) ^2 \left[ \xi K_0 K_1 - \frac{v^2 \xi^2 }{2 c^2} (K_1^2 -K_0^2)\right] , 
	\end{equation} 
	where $K_0$($K_1$) is the modified Bessel function of zeroth (first) order, as functions of $\xi =b_0 \varepsilon_x / \hbar \gamma v $ wher $\gamma=1/\sqrt{1-v^2/c^2}$ is the relativistic Lorentz factor and $v$ is the projectile velocity.
\end{comment}
%=======================================================
\subsection{\label{sec:norm} Justification for the selection of the multiplicative factor and the oscillator parameter}

%=======================================================
In the usual single-particle potential model approach, the neutron bound-state wave functions $u_{\ell_{0}j_{0}}^{J_{0}}$ are typically determined using a Woods-Saxon potential, with the radius parameter $r_0 = 1.15\sim 1.25$ fm,  diffuseness parameter $a = 0.5\sim 0.7$ fm, and a potential depth $V (I^{\pi}_c,nlj)$ adjusted to reproduce the experimental neutron separation energies, while also accounting for the excitation energy of the core.
The spectroscopic factor $S$ is usually obtained as the ratio of the measured partial cross-section to the calculated theoretical single-particle cross-section, i.e.,
$S=\sigma_{exp}/\sigma_{th}$. It can also be extracted from Coulomb breakup experiments as $S=B(E1)_{exp}/B(E1)_{th}$.
However, the extracted spectroscopic factor depends on the parameters defining the geometry of the Woods-Saxon potential. Specifically, changes in $r_0$ and $a$ influence the asymptotic normalization of the single particle wave function and its root mean square (rms) radius \cite{Pal03}.
For example, adopting a larger diffuseness parameter $a = 0.7$ fm yields a spectroscopic factor of 0.72(5) for the ground-state wave function of $^{15}$C, approximately 20\% smaller than the value obtained using $a =0.50$ fm, which gives 0.91(6) \cite{Nak09}. 
Similarly for $^{11}$Be, changing the geometry from $r_0=1.25$ and $a=0.7$ to $r_0=1.15$ and $a=0.5$ increases the spectroscopic factor from 0.61(5) to 0.74(6) \cite{Pal03}.
Thus, the value of $\left\langle r_{cv}^2 \right\rangle$ depends on the potential geometry. The spectroscopic factor can then be expressed as:
	\begin{equation} \label{eq:Br3}
		S =\frac{B(E1)_{\text{exp}}}{B(E1)_{\text{th}}}=\frac{\left\langle r_{cv}^2 \right\rangle_{\text{exp}}}{\left\langle r_{cv}^2 \right\rangle_{\text{th}}}
	\end{equation}	
	where the bound-state radial wave function is replaced by the overlap function $I_{\ell_{0}j_{0}}^{J_{}}=\sqrt{S_{\ell_{0} j_{0}}}u_{\ell_{0}j_{0}}^{J_{0}}$.			

	By selecting a harmonic oscillator wave function as the relative wave function for a weakly-bound system, we replace a wave function characterized by a long exponential tail with one that exhibits rapid Gaussian decay. Consequently, this choice leads to an underestimation of the nuclear radius and the $E1$ matrix elements, where
	\begin{equation} \label{eq:Br5}
		B(E1) =\frac{3}{4\pi}(Z_{\rm eff}^{(1)}e)^{2}\left\langle r_{R_{n_0 \ell_0}}^2 \right\rangle,
		%=\frac{3}{4\pi}(Z_{\rm eff}^{(1)}e)^{2}  \left\langle R_{0}^2 \right\rangle  \,,
	\end{equation}
	where  $\left\langle r_{R_{n_0 \ell_0}}^2 \right\rangle = R_0^2$ for all $R_{n_0 \ell_0}$ functions. To obtain the correct $\sqrt{\langle r^{2}_{cv}\rangle}$, we apply a multiplicative renormalization factor defined as
	\begin{equation}\label{eq:C02}
		C_0^2=\frac{B(E1)_{\text{exp}}}{B(E1)_{\text{th}}}=\frac{\left\langle r_{cv}^2\right\rangle_{\text{exp}}}{\left\langle r_{cv}^2 \right\rangle_{\text{th}}}=\left[ \frac{r_{cv}^{rms}}{R_0}\right]^2
	\end{equation}
	where $C_0^2 [\alpha(I_c^{\pi},n_0 \ell_0 j_0)]^2$ can act as a spectroscopic factor.
Therefore, Eq. (\ref{eq:dBl2}) becomes
	\begin{align}\label{eq:dB1}
		\frac{dB(E1)}{d\varepsilon} &=C_0^2\frac{3}{4 \pi}(Z_{\rm eff}^{(1)}e)^{2} \sum_{n_0 \ell_0 j_0} [\alpha(I_c^{\pi},n_0 \ell_0 j_0)]^2 \nonumber\\ &
		\times\sum_{\ell} \langle \ell_{0} 01 0|\ell 0 \rangle^{2} 
		\ \Bigg\vert   \int_{0}^{\infty}dr\
		r\ u_{\varepsilon \ell j}^{J}\left(  r\right)  \
		u_{\ell_{0}j_{0}}^{J_{0}}\left(  r\right)\Bigg\vert^{2}. 
	\end{align}
	Finally, the corresponding sum rule (Eq.~\ref{eq:Sumrule1B}) and cross-section calculations (Eq.~\ref{eq:dsde2}), must both incorporate $C_0^2$. To determine $C_0^2$, we derive an expression for $r_{cv}^{rms}$ in Sec.~\ref{sec:rcv}.
%If we interpret this as an effective wave function for the system, we must renormalise the operators accordingly. It seems somewhat plausible that this can be achieved by a multiplicative renormalisation of the radial coordinate in such operators, by multiplying each power of $r$ with a parameter $C_0$. This can, in turn, be extracted from the root-mean-square radius of the halo nucleus, or more accurately from the value of $\sqrt{\langle r^{2}_{cv}\rangle}$ as shown in Fig.~\ref{fig:coordinate}). Since it is first order in $r$, the naive single particle $E1$ operator is multiplied by $C_0$, given by
\begin{figure}[h]
	\centering
	\includegraphics[width=0.47\textwidth]{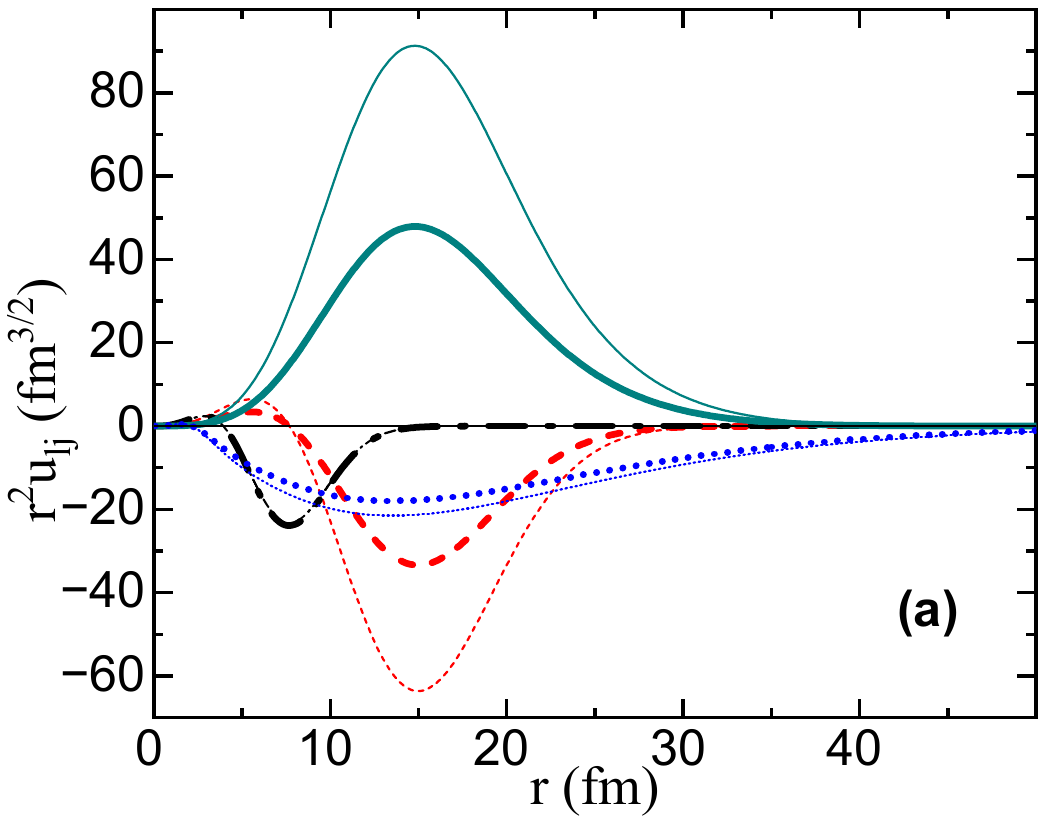}
	\includegraphics[width=0.47\textwidth]{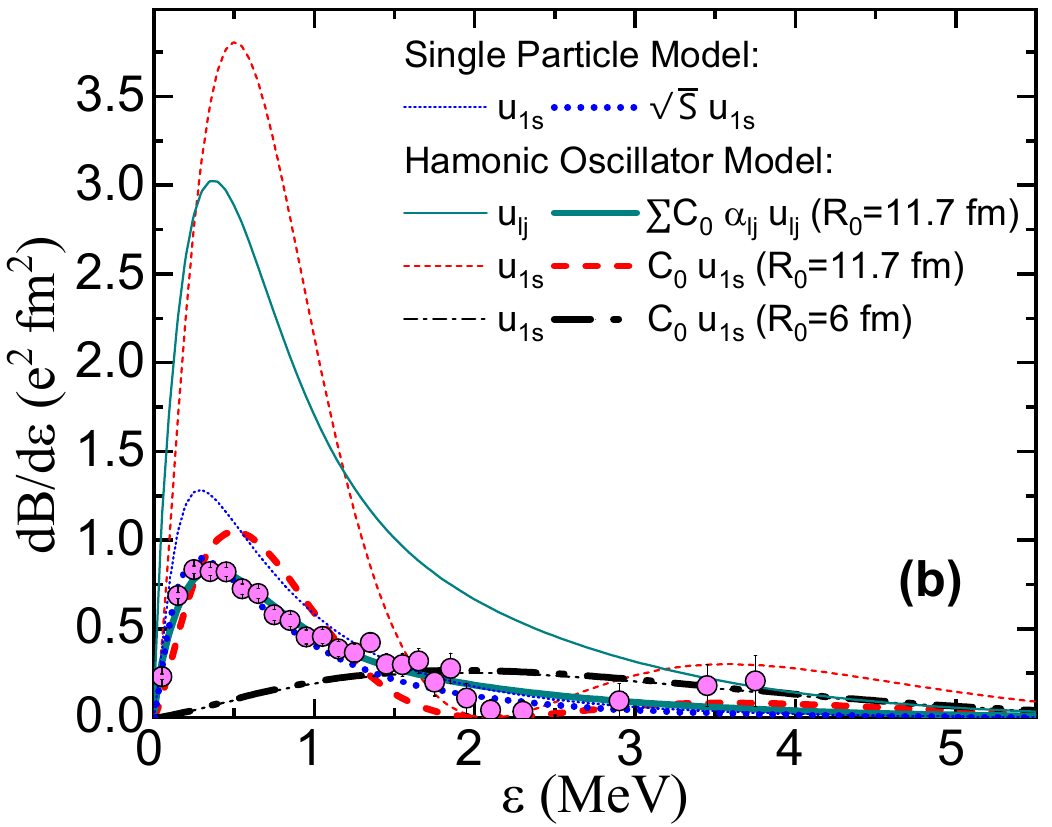}
	\caption{(a) Radial wave functions for $^{11}$Be, multiplied by $r^2$, as a function of the radial coordinate $r$. The dashed red lines represent $r^2 u_{\ell_0 j_0}^{J_0}(r)$ for the HO wave function of the $1s$ state, while the solid purple lines correspond to the sum of $s$, $p$, and $d$ HO states after fitting the dipole response function data, with $R_0 = 11.7$ fm. The dash-dotted black lines represent $r^2 u_{\ell_0 j_0}^{J_0}(r)$ for the HO wave function of the $1s$ state using $R_0 = 6$ fm. The dotted blue lines correspond to the single-particle wave function from Ref. \cite{Pal03}. The thin and thick lines represent the $r^2 u_{\ell_0 j_0}^{J_0}$ of HO wave functions before and after applying the multiplication factor $C_0$, with similar conventions for the single-particle model but using the spectroscopic amplitudes $\sqrt{S}$.		
		(b) The same as (a), but for $\frac{dB(E1)}{d\varepsilon}$.}\label{fig:clj}
\end{figure}

Since $u_{\varepsilon \ell j}^{J}(r)\sim r j_{\ell}(kr)$, the integrand in Eq.~(\ref{eq:dB1}) is given by the product of $r^2 u_{\ell_{0}j_{0}}^{J_{0}}(r)$ and $j_{1}(kr)$.
Figure \ref{fig:clj}(a) presents the values of $r^2 u_{\ell_{0}j_{0}}^{J_{0}}(r)$ for four different wave functions of $^{11}$Be: the $1s$ single-particle wave function calculated using a potential with geometry parameters $r = 1.25$ fm and $a = 0.7$ fm), the $1s$ HO wave function using $R_0 = 6$ fm, $1s$ HO wave function and the best-fit linear combination of $0s$, $0p$, $0d$, and $1s$ HO states that reproduces the dipole response function data from Ref. \cite{Fuk04}, with the oscillator parameter $R_0 = 11.72$ fm as determined by Eq. (\ref{eq:R0}). Further details regarding this fitting procedure are provided in Sec. \ref{sec:results}. 
%For the HO wave function, we adopt the oscillator parameter $R_0 = 11.72$ fm, as determined by Eq. (\ref{eq:R0}).

The thin lines in Fig. \ref{fig:clj} (a) represent $r^2 u_{\ell_{0}j_{0}}^{J_{0}}$ prior to applying the spectroscopic factor for the single-particle wave functions and the multiplicative factor $C_0$ for the HO wave functions. The thick lines correspond to the calculations after incorporating these factors. The area under the thick curves should be similar and reproduce the experimental value of $r_{cv}$ fm \cite{Fuk04}. 

The peak of $r^2 u_{\ell_{0}j_{0}}^{J_{0}}(r)$ for the HO wave functions is located at approximately $14$~fm, which closely aligns with the peak of the single-particle wave function. This agreement supports our choice of the oscillator parameter $R_0$ in Eq. (\ref{eq:R0}). In contrast, choosing $R_0$ based on the core-valence distance, as proposed by COSMA \cite{Zhu93}, approximately $6$~fm, inferred from the $^{11}$Be experiments, would shift the peak to around 7.5 fm, as indicated by the dashed-dotted lines in Fig. \ref{fig:clj} (a). A similar comparison was performed for $^{13}$C in Ref. \cite{Ots94}. The same issue arises in the calculation of $\frac{dB(E1)}{d\varepsilon}$, where choosing $R_0 = 6$ fm results in a peak that is significantly shifted from the first peak observed in the data.

The corresponding $\frac{dB(E1)}{d\varepsilon}$ values are presented in Fig. \ref{fig:clj}(b), highlighting the necessity of applying a spectroscopic factor for the single-particle wave function and a multiplicative factor for the HO wave function to accurately fit the $\frac{dB(E1)}{d\varepsilon}$ data and extract the $r_{cv}$ value. Additionally, the areas under the thick curves should yield the same integrated $B(E1)$ value.

The $1s$ HO wave function alone is insufficient to reproduce either the peak or the tail of the $\frac{dB(E1)}{d\varepsilon}$ data. However, a suitable combination of different states can achieve this, as detailed in the fitting procedure described in Sec. \ref{sec:results}.
 
The multiplicative renormalisation with powers of $C_0$ that we have applied to $\frac{dB(E1)}{d\varepsilon}$ could also be tested in the calculation of higher multipolarities, such as $\frac{dB(E\lambda)}{d\varepsilon}$. 
The natural assumption is that these would be renormalised by a factor $C_0^{2\lambda}$. However, due to the phenomenological nature of our method it is not immediately clear that this is an effective approach. To determine its validity would require comparison to detailed microscopic calculations, which falls outside the scope of this work.

%================================
\subsection{\label{sec:rcv} Estimating the distance between the Core and valence clusters}
%================================
\begin{figure}[t]
	\centering
	%trim=left botm right top
	\includegraphics[clip, trim=6.5cm 2cm 9.5cm 2cm, width=\columnwidth]{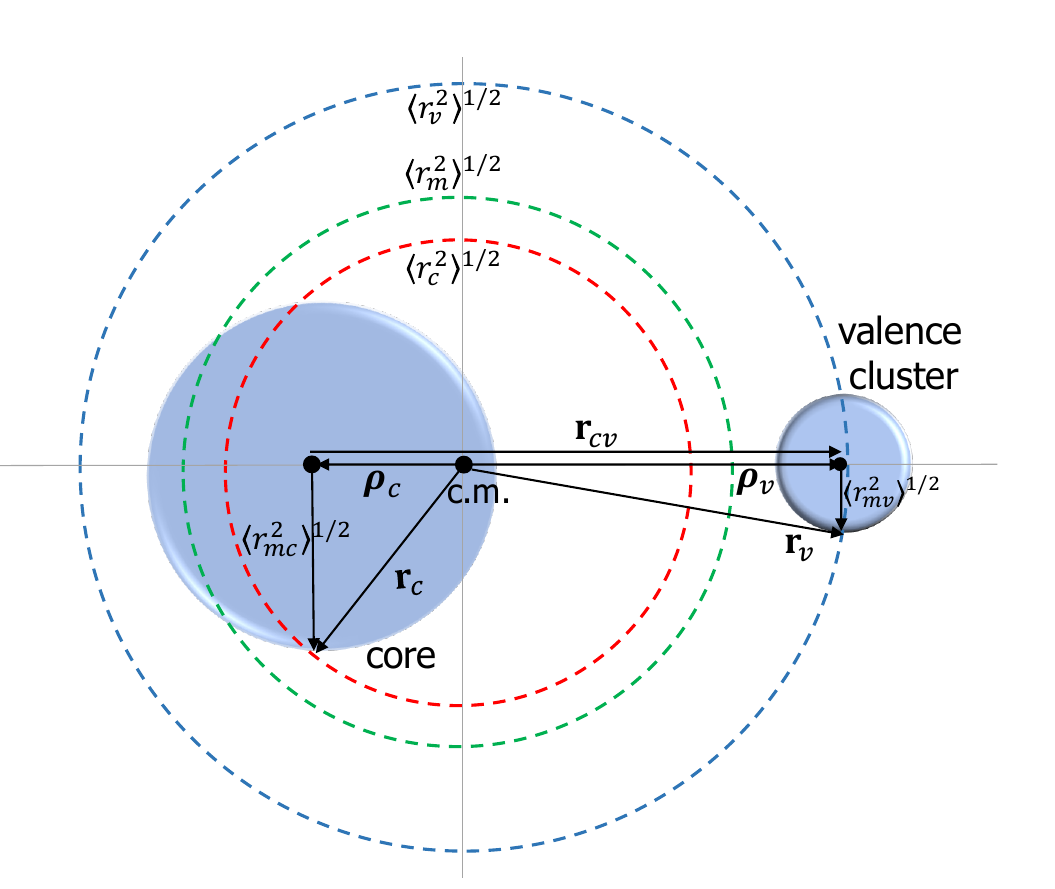}
	\caption{Coordinates of the two-body structure of the halo nucleus and for details refer to the main text.}
	\label{fig:coordinate}
\end{figure}
The mean distance from the centre of the core to the valence neutron(s) in one- and two-neutron halo nuclei, $r_{cv}^{rms}$, can be determined from the charge radius, the matter radius, and the $B(E1)$ value from the Coulomb dissociation measurement, respectively, as illustrated in Eq.~(\ref{eq:Br1}) \cite{Tanihata2013}.

Consider a halo nucleus with mass number $A$, consisting of a core with mass number $A_c$ and a cluster of valence neutrons with mass number $A_v$. Now we define
$r_{m}$, $r_{mc}$ and $r_{mv}$ as the internal coordinates relative to their centre of mass, inside of the halo nucleus, the core and valence cluster in free space, respectively. This allows us to extract the corresponding r.m.s.\ mass radii $r^{rms}_{m}$, $r^{rms}_{mc}$ and $r^{rms}_{mv}$, by weighing with the associated matter distributions (as illustrated in Fig.~\ref{fig:coordinate}). Inside the halo nucleus, the core and the valence cluster are displaced from their common centre of mass by the vectors $\vb*{\rho}_{c}$ and $\vb*{\rho}_{v}$ with $A_v \vb*{\rho}_{v} = -A_c \vb*{\rho}_{c}$,  so that we find the radii around this common centre of mass as
\begin{equation}\label{eq:rhocandv}
	\left\langle r_{i}^2 \right\rangle= \left\langle \rho_{i}^2 \right\rangle+\left\langle r_{mi}^2\right\rangle	\,,
\end{equation}
for $i=c,v$.

The matter radius of the halo nucleus is given by:
	\begin{equation}\label{eq:rm}
		A \left\langle r_{m}^2 \right\rangle=A_c \left\langle r_{c}^2 \right\rangle + A_v \left\langle  r_{v}^2	\right\rangle\,.
	\end{equation}
	By using Eq.~(\ref{eq:rhocandv}) in Eq.~(\ref{eq:rm}) we get
	\begin{equation}\label{eq:rm2}
		A \left\langle r_{m}^2	\right\rangle =  A_c \left\langle r_{mc}^2 \right\rangle + A_v \left\langle r_{mv}^2	\right\rangle +A_c \left\langle \rho_{c}^2 \right\rangle + A_v \left\langle \rho_{v}^2	\right\rangle.
	\end{equation}
	Using the relations $A_v \vb*{\rho}_{v} = -A_c \vb*{\rho}_{c}$ and $\vb*{r}_{cv}=-\vb*{\rho}_{c}+\vb*{\rho}_{v}$, with the magnitude ${r}_{cv}=\rho_{c}+\rho_{v}$, we derive:
	\begin{equation}\label{eq:rcv2}
		A_c \left\langle \rho_{c}^2 \right\rangle + A_v \left\langle \rho_{v}^2	\right\rangle =\frac{A_c A_v}{A} \left\langle r_{cv}^2	\right\rangle.
	\end{equation}
	Substituting this into Eq.~(\ref{eq:rm2}), we arrive at the expression:	
	\begin{equation}\label{eq:rm3}
		\left\langle r_{m}^2	\right\rangle =  \frac{A_c}{A} \left\langle r_{mc}^2 \right\rangle + \frac{A_v}{A} \left\langle r_{mv}^2	\right\rangle +\frac{A_c A_v}{A^2} \left\langle r_{cv}^2 \right\rangle 	
	\end{equation}
	which is equivalent to the expression in Ref.  \cite{Wal85,Mas09}.		
	The distance between the core and valence clusters is then given by:
	\begin{equation}\label{eq:rhocv}
		r^{rms}_{cv}=\sqrt{\frac{A}{A_c A_v} \left( A \left\langle r_{m}^2	\right\rangle -A_c \left\langle r_{mc}^2 \right\rangle -A_v \left\langle r_{mv}^2	\right\rangle \right) }	
	\end{equation}
	Similarly, analogous expressions for $\rho^{rms}_{c}$ and $\rho^{rms}_{v}$ can be derived from Eq.~(\ref{eq:rhocv}).
We should  mention that in case of one-neutron halo nuclei, using $A_v=1$ and neglecting the neutron radius, we will obtain the formula (3.11) of Tanihata et al. \cite{Tanihata2013}. For two-neutron halo nuclei, taking $A_v=2$ and using $r_{mv}=\frac{r_{nn}}{2}$, where $r_{nn}$ is the distance between the two valence neutrons, Eq. ($\ref{eq:rm3}$) is identical to the three-body formula \cite{Sagawa2015,Ber07}
\begin{equation}
	\label{eq:3b}
	\left\langle r_{m}^{2}\right\rangle=\frac{A_{c}}{A}\left\langle r_{m c}^{2}\right\rangle+\frac{2 A_{c}}{A^{2}}\left\langle r_{c v}^{2}\right\rangle+\frac{1}{2 A}\left\langle r_{nn}^{2}\right\rangle\,.
\end{equation}
For one-neutron halo nuclei, we use $r_{mv}=1\,\mathrm{fm}$.
For two-neutron halo nuclei described in our two-body model, we use $r_{mv}=\frac{r_{nn}}{2}$ with the experimental value of $r_{nn}$. 

The r.m.s. matter radii can be extracted from the measurement of reaction cross sections at intermediate to high energies using a Glauber model \cite{Ozawa2001,Tanihata2013}. They can also be obtained from high-energy elastic proton scattering \cite{Tanihata2013}.
%%%%%%%%%%%%%%%%%%%%%%%%%%%%%%%%%%%%%%%%%%%%%%%%%%%%%%%%%%%%%%%%%%%%%%%%%%%%%%%%%%%%%%%%%%%%%
\section{\label{sec:results} Results and discussions}	
\begin{table}[tbh]
	\caption{The combinations of the core spin-parity ($I^\pi_c)$ with different valence cluster orbits for the ground and first excited states of different halo nuclei. The $'*'$ denotes an excited state.}
	\vspace{0.2cm}
	\small
	\centering
	\begin{tabular}{ccccc}
		\hline\hline\noalign{\smallskip}
		Nucl. & Struc. & System     & core  & valence orbit   \\ 
		&core$+1n/2n$ & $J^\pi$  &$I_c^\pi$  &  $\ell / L$  \\ 
		\hline\noalign{\smallskip}
		%%%%%%%%%%%%%%%%%%%%%%%%%%%%%%%%%%%%%%%%%%%%%%%%%%%%%%%%%%%%%%%%%%%%%%%%%%%%%%%%%%%%%%
		$^{11}$Be&$^{10}$Be$+n$ &$1/2^+$  & $0^+$   &$s$ \\
		$^{11}$Be&$^{10}$Be$^*$$+n$ &$1/2^+$   &$2^+$,$1^-$,$0^+$    & $d$,$p$,$s$  \\
		$^{11}$Be$^*$&$^{10}$Be$+n$ &$1/2^-$ &$0^+$   &$p$ \\ [1.0ex]
		%	\toprule
		$^{15}$C&$^{14}$C$+n$& $1/2^+$ & $0^+$   &$s$ \\
		$^{15}$C&$^{14}$C$^*$$+n$&$1/2^+$   &$1^-$,$0^{+/-}$,$2^+$    & $p$,$s$,$d$  \\ 
		$^{15}$C$^*$&$^{14}$C$+n$ &$5/2^+$ &$0^+$   &$d$ \\[1.0ex]
		%	\toprule
		$^{6}$He&$^{4}$He$+2n$  & $0^+$ & $0^+$ &$S$ \\ 
		$^{6}$He$^*$&$^{4}$He$+2n$& $2^+$ & $0^+$ &$D$ \\ [1.0ex]
		%	\toprule
		$^{11}$Li&$^{9}$Li$+2n$  &$3/2^-$  & $3/2^-$ &$S$,$D$ \\
		$^{11}$Li&$^{9}$Li$^*$$+2n$	 &$3/2^-$  & $1/2^-$ &$D$\\
		\noalign{\smallskip}\hline\hline
		%%%%%%%%%%%%%%%%%%%%%%%%%%%%%%%%%%%%%%%%%%%%%%%%%%%%%%%%%%%%%%%%%%%%%%
	\end{tabular}
	\label{T1}
\end{table}
%%%%%%%%%%%%%%%%%%%%%%%%%%%%%%%%%%%%%%%%%%%%%%%%%%%%%%%%%%%%%%%%%%%%%%
We apply the formalism to two 1$n$-halo nuclei ($^{11}$Be and $^{15}$C) and two 2$n$-halo nuclei ($^{6}$He and $^{11}$Li). Both $^{11}$Be and $^{15}$C have $J^\pi=1/2^+$ ground states. There are two options to describe such states (i) the core ($^{10}$Be or $^{14}$C) is in the ground state ($I_c^\pi=0^+$) and the valence neutron is in the $0s/1s$ orbit, or (ii) the core is in one of the  $I_c^\pi=2^+,1^-,0^{+/-}$  excited states and the valence neutron is in either the $0d$, $0p$, or $0s/1s$ orbits. One can also combine the $1/2^-$ and $5/2^+$ excited states of $^{11}$Be and $^{15}$C  with a $0p$ or $0d$ neutron. 
In the case of $2n$ halos $^{6}$He and $^{11}$Li, the halo nuclei have the same ground state angular momentum as their core, $0^+$ and $3/2^-$, respectively. Thus the two neutrons in the valence cluster must be coupled to angular momentum zero ($0^+$), and the valence cluster must orbit the core in an  $S$ wave. To avoid confusion, we will use capital letters to denote orbits of the di-neutron-cluster state. In addition, we would like to mention that Eqs.~(\ref{eq:JM} \& \ref{eq:Yjm}) are still valid for the two-neutron halo systems in our model.
The $D$ wave can couple with the ground state and the first excited state of $^{9}$Li to produce the $3/2^-$ ground state of $^{11}$Li, though the spin of the excited state of $^{11}$Li remains unknown. For $^{6}$He, the $D$ wave contributes only to the first $2^+$ resonance and was included in calculations of breakup couplings in $^{6}$He scattering \cite{Rusek2000,Rusek2001}. 

Table \ref{T1} summarizes the possible angular momentum combinations for the wave functions of these nuclei, which include the low-energy excited states of the respective cores. By combining different orbits, as described in Eq.~(\ref{eq:dB1}), and ensuring the spectroscopic weights sum to unity, we can then adjust the weights to match experimental data.

Here, it is worth noting that in the absence of spin-orbit splitting, the computed $dB(E\lambda)/d\varepsilon$ values for $0d_{3/2}$ and $0d_{5/2}$ orbitals are identical. Consequently, we treat the $0d$ orbit of the neutron(s) in the ground state as a single entity. Similarly, we will also ignore the spin-orbit coupling for $p$ waves.
%%%%%%%%%%%%%%%%%%%%%%%%%%%%%%%%%%%%%%%%%%%%%%%%%%%%%%%%%%%%%%%%%%%%%%
\begin{table}[t]
	\centering
	\caption{\label{tab:rcv} The distance between the core and valence neutron(s) ($r^{rms}_{cv}$), the value of $R_0$, and the computed value of $C_0^2$ for different nuclei.}
	\begin{tabular}{cccccc}
		\hline\hline \noalign{\smallskip}
		System &$r^{rms}_{mc}$ &$r^{rms}_m$ &$r^{rms}_{cv}$ &$R_0$ & $C_0^2$  \\
		%		 \hline 
		core$+1n/2n$&fm &fm &fm &fm & \\ 
		\hline\noalign{\smallskip}
		$^{10}$Be+$n$ &2.39 \cite{Tanihata1985} &2.90 \cite{Kha96} &6.15 &11.72 &0.28 \\
		$^{14}$C+$n$  &2.43 \cite{Dobrovolsky2021} &2.60 \cite{Dobrovolsky2021} &4.36 &7.43 &0.34 \\
		$^{4}$He+$2n$ &1.57 \cite{Tanihata1988} &2.48 \cite{Tanihata1988} &3.87 &5.42  &0.55 \\
		$^{9}$Li+$2n$ &2.32 \cite{Tanihata1988} &3.12 \cite{Tanihata1988}&4.75 &9.84 &0.23 \\
		\noalign{\smallskip}\hline\hline           
	\end{tabular}
	%\end{ruledtabular}
\end{table}	

Determining $r_{cv}^{rms}$ values for the calculation of $C_0^2$ via Eq.~(\ref{eq:C02}) is straightforward using Eq.~(\ref{eq:rhocv}). The extracted value of $R_0$ depends on the separation energy $\mathcal{E}_0\equiv s_{n}$, see Eq.~(\ref{eq:R0}). Within a di-neutron model of $^{6}$He \cite{Rus01,Kee03,Rus03a,Rus04,Rus05,Mac04}, the binding energy of the di-neutron has been taken as $\mathcal{E}_0=s_{2n}=0.975$ and the relative wave function of two neutrons was assumed to be similar to that of two neutrons with zero relative energy which led to an unrealistic wave function for the di-neutron-$\alpha$ motion \cite{Mor07}. In their modified di-neutron model Moro \emph{et al} \cite{Mor07} adjusted the $2n$-core binding energy in $^{6}$He to reproduce the density distribution given by a realistic three-body model. They found that a separation energy of $1.6$ MeV can simulate the wave functions of realistic three-body calculations and gives a very good description of the elastic scattering data for several reactions induced by $^{6}$He.
In our work, we use a one-neutron separation energy of $\mathcal{E}_0=s_{n}=1.71$ MeV instead of the original reference value $\mathcal{E}_0=s_{2n}=0.975$ MeV, which is close to the value suggested by Moro \textit{et al} \cite{Mor07}. In a similarl way we use  $\mathcal{E}_0=s_{n}=0.396$~MeV  for $^{11}$Li.
The inputs of Eq.~(\ref{eq:rhocv}), the obtained values of $r^{rms}_{cv}$, $R_0$, and $C_0^2$ are listed in Table \ref{tab:rcv}. For instance, using the r.m.s. radii of $^{11}$Be and $^{10}$Be yields $r^{rms}_{cv}$ as $6.15$ fm. A similar procedure for $^{15}$C gives $r_{cv}=4.36$ fm. For the $^{6}$He and $^{11}$Li nuclei we get $3.87$ and $4.75$~fm respectively. 
In these calculations  we use $r_{mv}=1$~fm, the nucleon radius, for the one-neutron halo nuclei $^{11}$Be and $^{15}$C. For two-neutron halo nuclei we use $r_{mv}=r_{nn}/2$ using the experimental values of $r_{nn}$,  $3.93$~fm  for $^{6}$He  \cite{Tanihata2013} and $6.6$~fm for $^{11}$Li \cite{Marques2000}. It is clear that the obtained $r^{rms}_{cv}$ are in good agreement with those extracted from the experimental $B(E1)$ values as listed in table \ref{tab:dBresults}.
%%%%%%%%%%%%%%%%%%%%%%%%%%%%%%%%%%%%%%%%%%%%%%%%%%%%%%%%%%%%%%%%%%%%%%
\begin{figure}[t]
	\centering
	\includegraphics[width=0.47\textwidth]{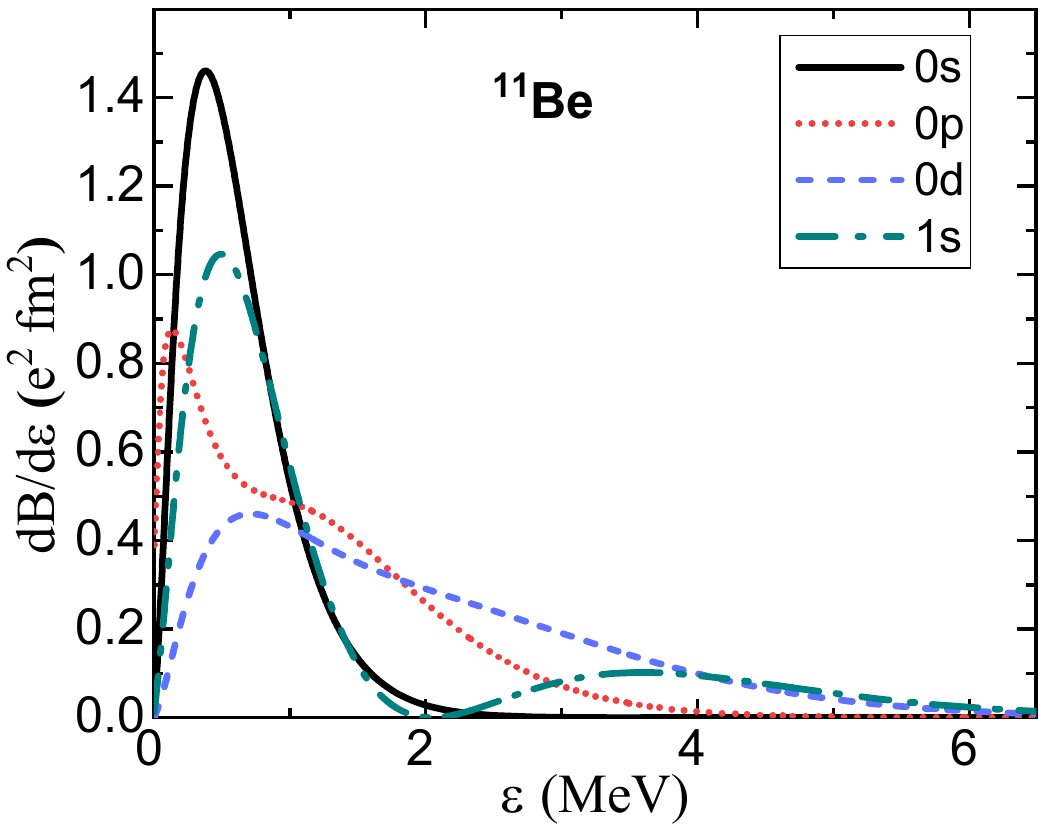}
	\caption{Dipole strength distribution $dB(E1)/d\varepsilon$ for different choices of the ground state for $^{11}$Be.}\label{fig:dBspdf}
\end{figure}
\begin{table*}[tbh]
	\centering
	\caption{\label{tab:dBresults} Results for the integrated $B(E1)$ (non-energy weighted), the core-valence radius $r_{cv}^{rms}$, and sum rule $S_1$ for one and two-neutron halo nuclei. The first five columns detail the probabilities assigned to the different states with the corresponding $\chi^2$ of the best-fit combination that reproduce $dB(E1)/d\varepsilon$. The remaining columns compare our calculations with experimental results and the sum rule values.}
	%\begin{tabularx}{\linewidth}{@{}cccccccC{1}cC{1}cc@{}}
		\begin{tabularx}{\linewidth}{@{}cccccccC{1}cC{1}cc@{}}
		%\begin{tabularx}{cccccccccccc}
		\hline\hline \noalign{\smallskip}
		System &\multicolumn{5}{|c|}{weight $\alpha^2$} &\multicolumn{2}{c|}{$B(E1)$ ($e^2$fm$^{2}$)} & \multicolumn{2}{c|}{$r_{cv}^{rms}$ (fm)}  & \multicolumn{2}{c}{$S_1$ ($e^2$fm$^{2}$)} \\
		\noalign{\smallskip} \cline{2-6}\cline{7-8}\cline{9-10}\cline{11-12} \noalign{\smallskip}  
		&\multicolumn{1}{|c}{$0s/0S$}&$0p$ &$0d/0D$ &$1s/1S$ &$\chi^2$ &\multicolumn{1}{|c}{Calc.} &exp. & \multicolumn{1}{|c}{Calc.}&  exp.  &\multicolumn{1}{|c}{Calc.} &Eq.(\ref{eq:Sumrule1}) \\
		%Proj.  & $R_0$  & $C_0$ &$0s$ &$0d$ &$B(E1)$\footnotemark[1]  &exp. $B(E1)$\footnotemark[2] & $r_{v}$\footnotemark[1] &  exp. $r_{v}$ &$S_1$ &$S_1$(\ref{eq:Sumrule1}) \\
		% \hline
		% &fm & & & &$e^2$fm$^{2}$ &$e^2$fm$^{2}$ &fm   &fm  &$e^2$fm$^{2}$ &$e^2$fm$^{2}$ \\ 
		\noalign{\smallskip} \hline  \noalign{\smallskip}
		\multirow{3}{*}{$^{10}$Be+$n$}&0.35&0.29&0.36 &0.0 &2.91 &\multirow{3}{*}{1.19} &\multirow{3}{\hsize}{\centering 0.90(6)\cite{Pal03}, 1.3(3)\cite{Nak94}, 1.05(6)\cite{Fuk04}}  & \multirow{3}{*}{6.15} &\multirow{3}{\hsize}{\centering 6.4(7)\cite{Nak94}, 5.7(4)\cite{Pal03}, 5.77(16), 6.1(5) \cite{Fuk04}} &2.10 & \multirow{3}{*}{2.18}\\
		&0.45& &0.55  & &9.52 &  &  &   & &2.23 &  \\
		&0.30&0.70&  & &27.4 &  &  &  & &3.73 &  \\
		\hline  \noalign{\smallskip}
		\multirow{3}{*}{$^{14}$C+$n$}&0.64&0.31&0.05  &0.0 &3.13 &\multirow{3}{*}{0.73} &\multirow{3}{\hsize}{\centering 0.53(5), 0.77(7) \cite{Nak09}}  & \multirow{3}{*}{4.36} &\multirow{3}{\hsize}{\centering 4.5(2) \cite{Nak09}} &2.38 & \multirow{3}{*}{2.55}\\
		&0.75&&0.25  & &6.02 &  &  &   & &2.54 &  \\
		&0.62&0.38&  & &4.15 &  &  &  & &2.32 &  \\
		\hline  \noalign{\smallskip}
		%$^{4}$He+$2n$ &5.24  &0.53   &0.38&&0.62  &1.32 & 1.2(2) \cite{Aum99}, 1.6(2) \cite{Sun21} &3.53   &3.36(39)\cite{Aum99}, 3.9(2)\cite{Sun21}&6.03&4.95 \\
		$^{4}$He+$2n$  &0.38&&0.62  &0.0 &10.4 &1.58 & 1.2(2) \cite{Aum99}, 1.6(2) \cite{Sun21} &3.86   &3.36(39)\cite{Aum99}, 3.9(2)\cite{Sun21}&10.30&4.95 \\
		%&0.30&0.70&  & &8000 &  & &  & &5.44 &  \\
		\hline  \noalign{\smallskip}
		\multirow{2}{*}{$^{9}$Li+$2n$} &0.21&&0.67  &0.11 &2.19 &\multirow{2}{*}{1.60} &\multirow{2}{\hsize}{1.42(18), 1.78(22) \cite{Nak06}} &\multirow{2}{*}{4.75} &\multirow{2}{\hsize}{5.01(32) \cite{Nak06}, 6.2(5) \cite{San06}}&2.68 &\multirow{2}{*}{2.70} \\
		&0.28&&0.72  & &2.52 &  & &  & &2.64 &  \\
		%&0.19&0.82&  & &161 &  & &  & &5.44 &  \\
		\noalign{\smallskip} \hline\hline       
	\end{tabularx}
\end{table*}
%%%%%%%%%%%%%%%%%%%%%%%%%%%%%%%%%%%%%%%%%%%%%%%%%%%%%%%%%%%%%%%%%%%%%%%%%%%%%%%%%%%%%%%%%%%%%%%%%%%%%%%%%%%%%%%%
\begin{figure*}[tbh]
	\centering
	\includegraphics[width=0.8\textwidth]{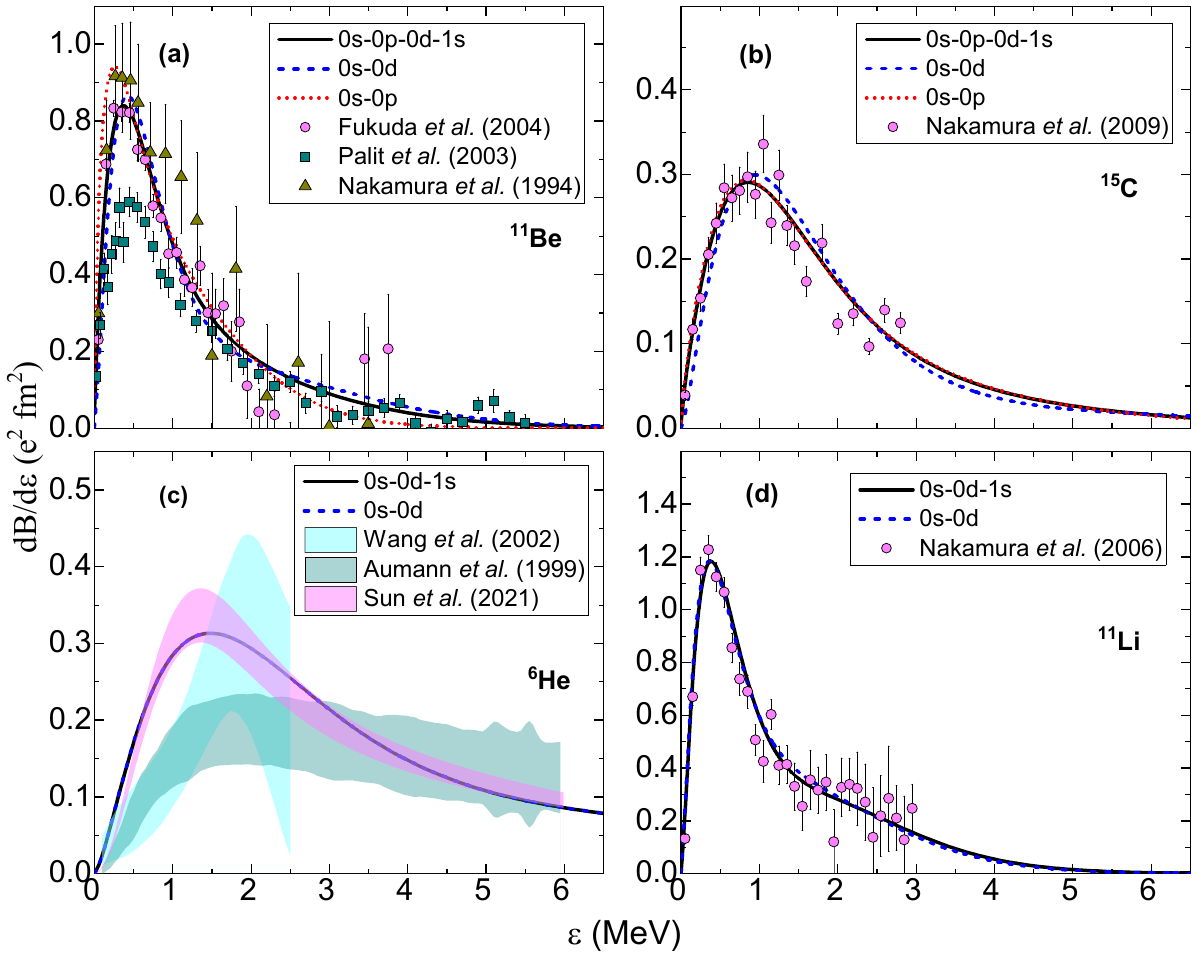}
	\caption{Dipole strength distribution $dB(E1)/d\varepsilon$ for one and two-neutron halo nuclei. The lines represent the calculations using different combinations of core-neutron cluster wave functions; refer to Table \ref{tab:dBresults} for detailed information on these combinations. The calculated $dB(E1)/d\varepsilon$ are compared with the experimental data.  (a) shows $^{11}$Be and data from Nakamura \textit{et al.} at $72$\,MeV/A \cite{Nak94}, Palit \textit{et al.} at $520$\,MeV/A \cite{Pal03}, and Fukuda \textit{et al.} at $69$\,MeV/A \cite{Fuk04}; (b) shows $^{15}$C and data from Nakamura \textit{et al.} at $68$\,MeV/A \cite{Nak09}; (c) gives results for $^{6}$He with data from Aumann \textit{et al.} at $240$\,MeV/A \cite{Aum99}, Wang \textit{et al.} at $23.9$\,MeV/A \cite{Wan02}, and Sun \textit{et al.} at $70$\,MeV/A \cite{Sun21}; (d) shows $^{11}$Li with data from Nakamura \textit{et al.} at $70$\,MeV/A \cite{Nak06}.}\label{fig:dBdE}
\end{figure*}
%%%%%%%%%%%%%%%%%%%%%%%%%%%%%%%%%%%%

In our framework, 
the systems are modeled as a two-body structure with a valence cluster of one or two valence neutrons coupled to an inert core.	
%	Like COSM model \cite{Suz88}, the excitation of the valence cluster does not lead to spurious center of mass excitations and no matter how highly the valence nucleons are excited. 
%	This is realized by introducing radius vectors of the valence nucleons relative to the core, not to the system center-of-mass.
%	Like COSMA \cite{Zhu93}, the $0s$ state of the valence neutron or cluster here is not forbidden by the Pauli principle because it has rather different oscillator radius than of the core $0s$ state hence they belong to different full sets.
The oscillator parameters ($b_{0s}$) of the $0s$ states for the valence neutron or clusters in the considered systems (e.g $\sim 8$ fm for $^{11}$Li) are significantly larger than the $0s$ states of their cores ($\sim 2$ fm for $^{9}$Li), as evident from Table \ref{tab:rcv}. This substantial difference allows us to assume that the $0s$ state of the valence neutron or cluster is not influenced by the Pauli exclusion principle. This interpretation aligns with the one made in the COSMA model \cite{Zhu93}.

We want to point out that a $s$-state gives rise to a low-energy peak in $dB(E\lambda)/d\varepsilon$, a $p$ wave moves the peak to the left and contributes to the medium-energy region but not as much as a $d$-state which has a long energy tail. 
%We thus prefer a combination of $s$ and $d$ waves. 
This can be seen in Fig.~\ref{fig:dBspdf} for the case of $^{11}$Be. All the calculations shown in Fig.~\ref{fig:dBspdf} have the same integrated $B(E1)$ value and identical $r_{cv}^{rms}$.
%%%%%%%%%%%%%%%%%%%%%%%%%%%%%%%%%%%%%%%%%%%%%%%%%%%%%%%%%%%%%%%%%%%

The calculation of $dB(E1)/d\varepsilon$ involves utilizing Eq.~(\ref{eq:dB1}), where we search for weights $\alpha^2$ for different states that best fit the experimental data. 
The $\chi ^2$ statistic is defined as:
\begin{equation}
\label{eq:chi2}
\chi ^2 = \frac{1}{N}\sum_{k=1}^{N}\left[\frac{B_{\mathrm{th}}(\varepsilon_k)-B_{\mathrm{ex}}(\varepsilon_k)}
{\Delta B_{\mathrm{ex}}(\varepsilon_k)}\right]^{2},
\end{equation}
where $B_{\mathrm{th}}(\varepsilon_k)$ and $B_{\mathrm{ex}}(\varepsilon_k)$ are the theoretical and experimental dipole response functions at the relative energy $\varepsilon_k$, respectively. $\Delta B_{\mathrm{ex}}(\varepsilon_k)$ is the experimental error and $N$ is the number of data points.
%Sometimes, the minimum $\chi ^2$ value does not always correspond to a better visual result. In such cases, a subjective assessment of the fit quality (based on visual inspection) may hold greater significance than the $\chi ^2$ value.
For the fitting procedure, we use experimental data from: Fukuda \textit{et al.} data \cite{Fuk04} for $^{11}$Be, Nakamura \textit{et al.} data \cite{Nak09} for $^{15}$C, Sun \textit{et al.} data \cite{Sun21} for $^{6}$He, and Nakamura \textit{et al.} data \cite{Nak06} for $^{11}$Li.
Initially, we explored combinations involving $0s/0S$, $0p$, $0d/0D$, and $1s/1S$ states but found inadequate agreement with the data, with the $1s/1S$ state either not contributing, or making a small contribution (for $^{11}$Li), as shown in Table~\ref{tab:dBresults}.  
It should be noted that the values of $\chi^2$, defined as an error-weighted sum, are rather large: this is due to small deviations at low energy, where data often has very small statistical errors. This is especially the case for $^6$He, where we slightly overestimate the strength distribution below $0.1\,\mathrm{MeV}$.

For $^{11}$Be, the first and second excited states of $^{10}$Be, both $J^\pi=2^+$, can couple with the $0d$ orbit, while the $0p$ contribution would arise from coupling to a higher-energy $1^-$  state, as shown in Table \ref{T1}. This explains the larger contribution of $0d$ compared to $0p$ in the fitting. We achieve a good fit using the $0s$ and $0d$ combination, but not with the $0s$ and $0p$ combination, as shown by the $\chi^2$ values in Table~\ref{tab:dBresults}. Whereas, for the $^{15}$C fitting, the roles of $0p$ and $0d$ are reversed. The $0p$ contribution is more significant because it couples with the first excited state of the core ($^{14}$C), $1^-$, and the higher $0^-$ state. In contrast, the $0d$ orbit can only couple with the fifth excited state, $2^+$, which explains its smaller contribution. 

For $^{6}$He, we find that the data is fitted well with a combination of $0S$ and $0D$, where the $0D$ wave can only correspond to a contribution of the $2^+$ excited state of $^{6}$He. For $^{11}$Li, we find that the combination of the $0S$ and $0D$ orbits fits the $^{11}$Li data well. It is important to note that the $0D$ orbit has a contribution from the coupling of both the ground and first excited states of the $^9$Li core. In summary, we obtain a good fit to the data with a single free parameter, $0s/0S$, in combination with orbits that couple to the first excited states of the cores or nuclei itself: $0d/0D$ for $^{11}$Be, $^{6}$He, and $^{11}$Li, and $0p$ for $^{15}$C.

The calculated $dB(E1)/d\varepsilon$ for these four one- and two-neutron halo nuclei are shown in Fig.~\ref{fig:dBdE}, with the corresponding integrated $B(E1)$ values and $r_{cv}$ listed in Table \ref{tab:dBresults}. 
Remarkably, the $0s/0S$-$0d/0D$ or $0s$-$0p$ combinations consistently provide excellent agreements with experimental data across all nuclei. Furthermore, the extracted $B(E1)$ and $r_{cv}$ values closely match the experimental data. Moreover, the calculated $S_1$ from $dB(E1)/d\varepsilon$ aligns well with the theoretical cluster sum rule values described by Eq.~(\ref{eq:Sumrule1}).

Our calculations for the nuclei $^{11}$Be, $^{15}$C, $^{6}$He, and $^{11}$Li shown in panels (a), (b), (c), and (d) respectively, of Fig.~\ref{fig:dBdE}, yield excellent results that agree well with the experimental data. 
Specifically, for $^{11}$Be, our model aligns with the $dB(E1)/d\varepsilon$ data \cite{Fuk04} through the fitting of weights in Table~\ref{tab:dBresults}. The resulting $B(E1)$ and $r_v$ values align closely with experimental observations. Despite variations in extracted $S_1$ values, our model consistently produces values near the total cluster sum rule, affirming its robustness. 
Similarly, for $^{15}$C, our approach agrees well with the experimental data, with calculated $S_1$ values mirroring theoretical cluster sums. 
In $^{6}$He, our model accurately reproduces the peak position at 1.4 MeV and shape of $dB(E1)/d\varepsilon$ of the data \cite{Sun21}, with integrated $B(E1)$ strength and $r_v$ in line with experimental findings, demonstrating good agreement with previous results. Lastly, in $^{11}$Li, we find favourable fits to data, with integrated $B(E1)$ and $R_v$ values consistent with experimental reports, and $S_1$ values aligning well with theoretical cluster sums. These consistent successes across multiple nuclei underscore the efficacy and reliability of our model for dipole strength. 
Our findings indicate that the model's validation and the renormalization procedure are specifically tailored to aspects of the dipole strength and may not extend to other observables. To address this, we aim to utilize these HO wave functions in the near future to test their validity for transfer calculations.
%=======================================================
%\subsection{\label{sec:summary} cross sections}
%=======================================================
\begin{figure*}[tbh]
	\centering
	\includegraphics[width=0.8\textwidth]{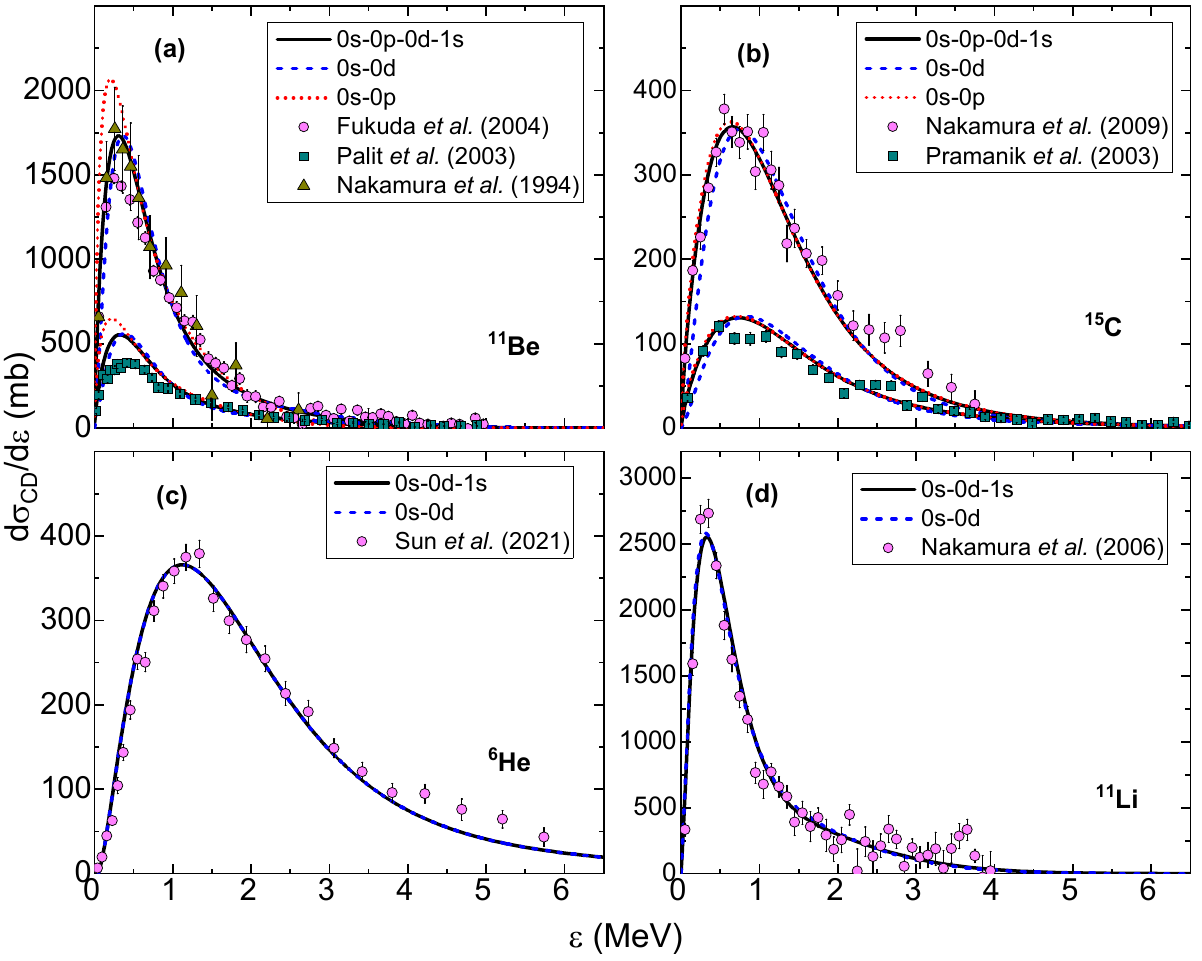}
	\caption{Coulomb Breakup cross sections ($d\sigma_{CD}(E1)/d\varepsilon$) as a function of the relative energy for 
		(a) $^{11}$Be+Pb in comparison with the data of Nakamura \textit{et al.} at $72$\,MeV/A \cite{Nak94}, Palit \textit{et al.} at $520$\,MeV/A \cite{Pal03}, and Fukuda \textit{et al.} at $69$\,MeV/A \cite{Fuk04}; 
		(b) $^{15}$C+Pb in comparison with the data from Nakamura \textit{et al.} at $68$\,MeV/A \cite{Nak09} and  Pramanik \textit{et al.} at $605$\,MeV/A \cite{Pra03};
		(c)$^{6}$He+Pb in comparison with the data from Sun \textit{et al.} at $70$\,MeV/A \cite{Sun21};
		(d)$^{11}$Li+Pb in comparison with the data from Nakamura \textit{et al.} at $70$\,MeV/A \cite{Nak06}.
		The lines represent the calculations using different combinations of core-neutron cluster wave functions; refer to Table \ref{tab:dBresults} for detailed information on these combinations. }\label{fig:dSdE}
\end{figure*}

The Coulomb dissociation cross section, calculated using Eq. \ref{eq:dsde2} and the $dB(E1)/d\varepsilon$ values presented in Table \ref{tab:dBresults}, is shown in Fig.~\ref{fig:dSdE} for the four halo nuclei studied previously interacting with a lead target at intermediate and higher energies. The calculated results show excellent agreement with the experimental data.
It is important to note that the data displayed in the figure represents the Coulomb breakup cross sections, i.e., the cross section with the nuclear contributions removed except for $^{15}$C+Pb at $68$\,MeV/A and $^{11}$Li+Pb at $70$\,MeV/A where there is no available nuclear breakup data. These nuclear contributions were estimated from measurements obtained with a carbon target, ensuring the results reflect purely electromagnetic dissociation by using 
\begin{equation}
	\label{eq:scale}
	\frac{d\sigma_{CD}}{d\varepsilon}=\frac{d\sigma}{d\varepsilon}(\mathrm{Pb})-\Gamma\frac{d\sigma}{d\varepsilon}(\mathrm{C})
\end{equation}
where $\Gamma$ is a scaling factor. Assuming the nuclear excitation is predominantly peripheral, $\Gamma$ can be taken as the ratio of the sum of the radii of the target and the projectile, which is approximately 1.8 \cite{Fuk04,Nak94}.
Alternative estimates for $\Gamma$, as well as other methods for excluding the nuclear contribution, are detailed in the references cited in the caption of Fig.~\ref{fig:dSdE}. 
%=======================================================
%
\section{\label{sec:summary} Summary and conclusion}
%
%=======================================================
In summary, a simple cluster shell-model approximation is presented to estimate the ground state wave function of halo nuclei and applied successfully to reproduce the soft $E1$ response function distribution for the 1$n$-halo nuclei $^{11}$Be and $^{15}$C and 2$n$-halo nuclei $^{6}$He and $^{11}$Li.	This is a very powerful and surprising result, in light of the very simple wavefunctions and the equally simple operator renormalisation used.  However, with only a single free parameter, the $sd/SD$ or $sp$ mixing, we are able to give a very good quantitative description of the experimental results.
In addition, the corresponding Coulomb dissociation cross section data are reproduced well.  
%Also, a new simplified expression for the distance between the two clusters of halo nuclei is presented. In general, this expression can be helpful in determining the distance between two clusters within a nucleus, for example $\alpha+t$ structure of $^{7}Li$ and $\alpha+d$ structure of $^{6}Li$, using the r.m.s. radius of the nucleus and its clusters as free nuclei.
In the future, we hope to give a more microscopic foundation for the phenomenological renormalisation used. We also intend to see whether a similar argument can be used effectively for higher multipolarities, heavier nuclei and at the proton dripline.

%%%%%%%%%%%%%%%%%%%%%%%%%%%%%%%%%%%%%%%%%%%%%%%%%%%%%%%%%%%%%%%%%%%

\section*{Acknowledgments}
This work was funded by the Council for At-Risk Academics (Cara) within the Cara Fellowship Programme \& partially supported by the British Academy within the British Academy/Cara/Leverhulme Researchers at Risk Research Support Grants Programme under grant number LTRSF/100141 (HMM) and by the UK Science and Technology Funding Council [grant numbers ST/V001116/1 and ST/Y000323/1] (JS, NRW, and DKS).	%%%%%%%%%%%%%%%%%%%%%%%%%%%%%%%%%%%%%%%%%%%%%%%%%%%%%%%%%%%%%%%%%%%%%%%%%
%	\newpage
\bibliography{ref}

\end{document}